# A Helical Model for the Compact Jet in 3C 345

W. Steffen[1,2], J. A. Zensus[3], T. P. Krichbaum[1], A. Witzel[1], and S. J. Qian[4]

[1] Max-Planck-Institut für Radioastronomie, Auf dem Hügel 69, 53121 Bonn, Germany
[2] Now at University of Manchester, Dept. of Physics and Astronomy, Manchester M13 9PL, UK
[3] National Radio Astronomy Observatory, 520 Edgemont Road, Charlottesville, VA 22903, USA
[4] Beijing Astronomical Observatory, Chinese Academy of Sciences, Beijing 100080, China

**Abstract.**
We present a simple model for the apparent superluminal motion along curved trajectories of features observed in the compact radio jet of the quasar 3C 345. The model is inspired by the magneto-hydrodynamic approach of Camenzind (1986), and assumes a conical geometry of the jet within about three milliarcseconds of the radio core. Simultaneous conservation of three out of four physical parameters is used to derive analytic solutions for the equation of motion of the idealized jet features. These are the jet's kinetic energy, angular momentum, and momentum along the jet axis. Conservation of angular momentum is required to fit the observed trajectories at distances larger than 3 mas, in agreement with the kinematic predictions of Camenzind's model. The best-fit model variant preserves angular momentum, kinetic energy, and opening angle, with an inclination of the jet axis to the line of sight $\theta = 6.8°$, an opening angle $\psi = 0.95°$, and Lorentz factors $\gamma = 5.8$ and $\gamma = 4.6$ for components C4 and C5, respectively ($H_0 = 100 \text{ km s}^{-1} \text{ Mpc}^{-1}$). The model ignores turbulent small-scale motion of the jet components, and it does not take into account the overall bending of the jet beyond about 4 mas, restricting it's applicability to the inner few milliarcseconds from the core.

## 1. Introduction

Curved radio structures and bent trajectories of moving features are frequently found in radio jets on parsec and kiloparsec scales. In some cases, the bends are quasi-periodic, suggesting a helical nature of the underlying structure. Examples are filaments in M 87 (Biretta et al. 1983), an oscillating ridge line in 3C 273 (Zensus et al. 1988, Krichbaum & Witzel 1992), bending and acceleration in 3C 84 (Krichbaum et al. 1992 and 1993b), a bent ridge line in 1803+78 (Krichbaum et al. 1994, Steffen 1994) and 3C 380 (Kus 1993). Periodic optical variability, e.g., in 3C 345 (Schramm et al. 1993) and PKS 0420-014 (Wagner et al. 1995), and swings in the polarization angle of OJ 287 (Kikuchi et al. 1988) and 0917+624 (Quirrenbach et al. 1989, Qian et al. 1991) have also been attributed to motion in helical jets.

Theoretical models for such oscillating, bent structures include helical modes in hydrodynamic jets (Hardee 1987, Owen et al. 1989) or in magnetized jets (Königl & Choudhuri 1985). Camenzind (Camenzind 1986, Camenzind & Krockenberger 1992) developed a model for compact jets with bulk plasma acceleration along helical magnetic field lines, based on magnetized accretion disc winds. This model predicts helical motion of off-axis features in the jet; through conservation of angular momentum in a conical jet geometry the model predicts also the asymptotic dampening of the jet oscillations seen in some cases.

The quasar 3C 345 contains such a bent core-jet structure on parsec-scales, and also is one of the best-studied sources with apparent superluminal motion (Biretta et al. 1986; Zensus et al. 1995). Several moving features have been traced during their evolution; these 'components' separate from the core on apparently different curved trajectories, and undergo changes of superluminal speed and flux density. The curvature of the trajectories is strongest within about 2 mas from the core, where components C4 and C5 have been traced; C4 has made at least one complete oscillation around the mean jet axis. Newly detected features close to the core also appear to move on different curved paths (Zensus et al. 1995; Krichbaum et al. 1993a). At larger distances, the apparent paths merge and straighten out.

These general properties were reconciled by Qian et al. (1992) in a geometrical model that explains the trajectories of C4 and C5 in 3C 345 as motion along helical paths. They introduced smooth (parabolic) bending of the jet axis in the sky plane to account for the observed bending of the jet at core separations larger than about 3 mas. While this model gives a satisfactory fit to the component trajectories and also allows derivation of co-moving un-boosted source parameters

(Qian et al. 1995), it does not provide a physical explanation for the origin of the helical structures, and it does not explain the straightening of the jet at larger distances from the core.

In this paper, we discuss the trajectories of components in 3C 345 by implementing the kinematic aspects of the model of Camenzind (Camenzind & Krockenberger 1992). This model predicts helical structures as a natural consequence of twisting magnetic field lines that are 'frozen' in a wind from a rotating accretion disc. From conservation of angular momentum, the model also predicts that the curvatures cease at large distances from the core. We ignore the complication of overall jet bending and restrict ourselves to the inner 3 mas from the core. We further assume motion with constant Lorentz factor $\gamma$, constant specific angular momentum $L_z$, and a constant opening half-angle $\psi$ (in the following called "opening angle") of the jet. For generality, we consider other combinations of conservation quantities in addition to those of Camenzind's model. We also include the specific momentum along the jet axis $p_z$, which may be used to parameterize other observed jets or models (e.g., Hardee 1987).

Section 2 describes the major variants of our model and the analytical solutions to the basic equations of motion. In Section 3 we apply the model to 3C 345; the results are summarized in Section 4.

## 2. The Helical Models

The moving features in the milli-arcsecond jet of the quasar 3C 345 are modeled as emission regions whose centres of gravity are moving on 3-dimensionally bent trajectories (Fig. 1). The motion is determined by the conservation of the specific kinetic energy (equivalent to a constant Lorentz factor $\gamma$) of the plasma, the specific angular momentum $L_z$ with respect to the jet axis, the specific momentum component along the jet axis $p_z$, and the opening angle of the jet $\psi$ (Fig. 2). The jet axis is assumed to be a straight line and is oriented in the z-direction at an angle $\theta$ to the observer's line of sight. The conservation laws for kinetic energy (1), angular momentum (2), momentum (3) and jet opening angle (4) can be expressed as:

$$E_{kin} = \gamma m_0 c^2 = \text{const} \tag{1}$$

$$L_z = \frac{\gamma m_0 \omega_{(t)} r_{(t)}^2}{\gamma m_0} = \omega_{(t)} r_{(t)}^2 = \text{const} \tag{2}$$

$$p_z = \frac{\gamma m_0 v_z}{\gamma m_0} = v_z = \text{const} \tag{3}$$

$$\tan \psi = \text{const} \tag{4}$$

Throughout, we use the following notation:

| | |
|---|---|
| $\gamma$ | Lorentz factor of the plasmon |
| $\psi$ | opening half-angle of the jet |
| $\theta$ | inclination angle of the jet axis |
| $m_0$ | rest mass of the emitting plasmon |
| $r_{(t)}$ | distance from the jet axis |
| $t$ | time in the frame of rest of the quasar |
| $\omega$ | angular rotation velocity $(d\phi/dt)$ |
| $z, r, \phi$ | cylindrical coordinates (fig. 1) |
| $v$ | velocity in units of the speed of light |
| $v_z, v_r, v_\phi$ | velocity components in units of the speed of light |

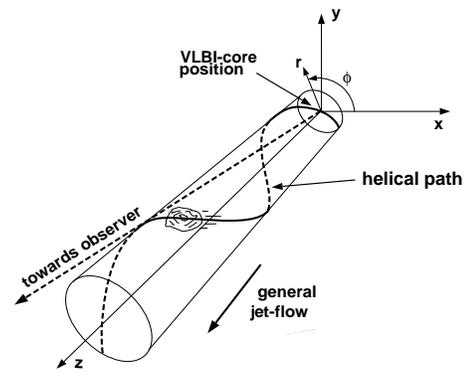

**Fig. 1.** Coordinate system used in the model calculations.

We now turn to the analytical solutions of the equations of motion, which we will use to calculate the position of the jet component as a function of time. We start the calculations at some initial point ($z = 0$) at a distance $r_0$ from the jet axis, with some initial angular velocity $\omega_0$ and Lorentz factor $\gamma_0$. Initial values are denoted by a subscript 'o'. The origin of the coordinate system is identified with the position of the VLBI-core of 3C 345 which is assumed to be stationary (Bartel et al. 1986).

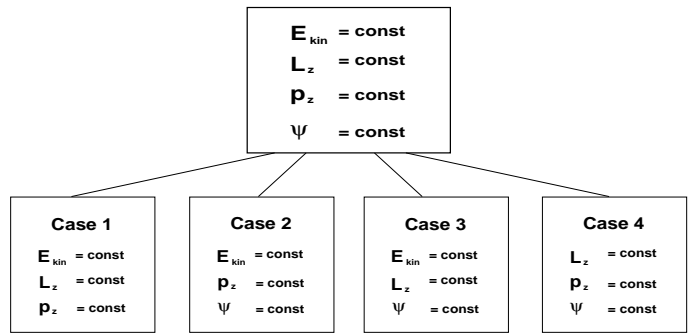

**Fig. 2.** Three out of four quantities are assumed to be constant during the motion of a jet component: the Lorentz factor $\gamma$, the specific angular momentum $L_z$ and the specific momentum along the jet axis $p_z$, and the opening angle of the jet $\psi$. Three of them are needed for every specific case.

### 2.1. Case 1

This model variant assumes the conservation of only mechanical quantities (momentum along the jet axis $p_z$, the kinetic energy $E_{kin}$, and the specific angular momentum $L_z$). There are no assumptions for the jet opening angle. Integration of the equations of motion (3, 1 & 2, in this order), leads to

$$z_{(t)} = v_z t + z_0 \tag{5}$$

$$r_{(t)} = \sqrt{\alpha v_z^2 t^2 + \delta v_z t + \zeta} \tag{6}$$

$$\phi_{(t)} = \phi_0 + \arctan \frac{\xi t + \eta}{L} - \arctan \frac{\eta}{L} \tag{7}$$

where the constants are $a = v^2/v_z^2 - 1$, $b = (2ar_0 - L^2/v_z^2)^{1/2}$,
$\zeta = r_0^2$, $\xi = \alpha v_z^2$, $\eta = (\xi r_0^2 - L^2)^{1/2}$.

Here and in the following sections the constant of integration $z_0$ is set to $z_0 = 0$.

Eqs. (5) & (6) show that the whole jet (or the envelope of all possible trajectories) is a rotational hyperboloid of increasing or decreasing opening angle (depending on the parameters $\alpha$, $\delta$, and $\zeta$). It follows from Equation (7) that for this case no complete revolution of the component around the jet axis is possible. Taking into account the properties of the arctan-function in the limit $t \to \infty$ it follows that $\phi_\infty - \phi_0 < \frac{\pi}{2}$. Thus only less than a quarter of a revolution can be achieved. The helix opens right away, and transforms into an almost straight line before undergoing a considerable bend. Therefore, this model variant is inadequate for a description of the oscillating motion of C4 in 3C 345 (fig. 7). Thus, we conclude that at least one of the quantities like the specific angular momentum, the Lorentz factor and the momentum component along the jet axis is *not* conserved by the moving emission regions in 3C 345.

### 2.2. Case 2

Case 2 is determined by the momentum along the jet axis $p_z$, the the kinetic energy $E_{kin}$ and the opening angle $\psi$.

Integration of the equations of motion (1, 3 & 4) leads to

$$z_{(t)} = v_z t \tag{8}$$
$$r_{(t)} = v_z t \tan\psi + r_0 \tag{9}$$
$$\phi_{(t)} = \phi_0 + \frac{\omega_0 r_0}{v_z \tan\psi} \ln \frac{r_{(t)}}{r_0}. \tag{10}$$

Figure 3 illustrates the general properties of the trajectory, i.e. the self-similarity and the constant pitch angle of the helix. These properties follow from computing the time derivative $\omega = d\phi/dt$ of Equation (10). It is found that the azimuthal velocity component $v_\phi = r\omega$ is constant in this case. From this result and the assumption of a constant momentum along the jet axis (3), we also find the pitch angle $\epsilon$ of the helix to be conserved along the jet:

$$\tan \epsilon = \frac{dz}{r\,d\phi} = \frac{v_z}{r\omega} = \text{const}, \tag{11}$$

This and the assumption of a constant opening angle cause an increasing "wavelength" of the helix along the jet (Fig. 3). This can be seen from calculating the wavelength $\lambda_{(z)} := z_{(\phi+2\pi)} - z_{(\phi)}$ and the period $P_{(t)} := t_{(\phi+2\pi)} - t_{(\phi)}$:

$$\lambda_{(z)} = \left(z + \frac{r_0}{\tan\psi}\right)\left(e^{2\pi\kappa} - 1\right) \tag{12}$$
$$P_{(t)} = \left(t + \frac{r_0}{v_z \tan\psi}\right)\left(e^{2\pi\kappa} - 1\right) \tag{13}$$

where $\kappa = (v_z \tan\psi)/(\omega_0 r_0)$. $\lambda_{(z)}$ and $P_{(t)}$ are linear in their variables and the parameters do not vary along the path. The trajectory therefore is self-similar in space and time, i.e. it can be scaled linearly in all four coordinates separately. In principle this model variant can explain helical motion, but only in a range where the helical amplitude is not dampened.
This case is equivalent to the kinematics of the hydrodynamic isothermal helical model presented by Hardee (1987). His set of equations $r = z\Psi + r_0$, $\phi = \phi_0 \pm \frac{2\pi r_0}{\Psi \lambda_0} \ln(r/r_0)$ (eq. 19a, adapted to our notation) and the condition of a constant opening angle, are similar to our scenario. This can be shown by setting $\Psi = \tan\psi$ (or $\Psi = \psi$ for small opening angles $\psi$) and expanding the exponential in Equation (12) into a power series. Truncation after two terms yields an initial wavelength $\lambda_0 \sim 2\pi v_z/\omega_0$ using $z = z_0 = 0$. Inserting this result into our equation for $\psi_{(t)}$ yields exactly Hardee's Equation 19a. The main difference to our simplified model is, that in the hydrodynamical model the jet as a whole is bent. Therefore, it is not easy to explain the different paths taken by closely following jet components (like C4 and C5), since the overall jet bending is assumed to change slowly compared to the motion of the individual components. For an application to 3C 345 this case is inadequate, since the oscillation of the component path continues up to infinite distances.

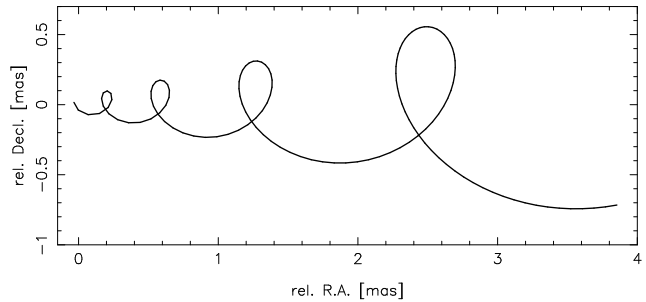

**Fig. 3.** A typical trajectory showing the properties of Case 2. The trajectory is self-similar in space and time.

### 2.3. Case 3

This case is based on the conservation of $E_{kin}$, $L_z$, and $\psi$ along the trajectory and represents an approximation to the kinematics expected from the model by Camenzind (Camenzind 1986, Camenzind & Krockenberger 1992).

The solution of the corresponding system of Equations (1, 2 & 4) again provides the position as a function of time:

$$r_{(t)} = \sqrt{(at+b)^2 + c^2} \tag{14}$$
$$z_{(t)} = \frac{r_{(t)} - r_0}{\tan\psi} \tag{15}$$
$$\phi_{(t)} = \phi_0 + \frac{1}{\sin\psi}\left[\arctan\frac{at+b}{c} - \arctan\frac{b}{c}\right] \tag{16}$$

where $a = v \sin\psi$, $b = \sqrt{r_0^2 - \frac{L^2}{v^2}}$, $c = L/v$. Note that expression (16) is similar to Equation (7) except for the factor $1/\sin\psi$. This counteracts the limit introduced by the *arctan* function on the phase angle $\phi_{(t)}$. Therefore, in this model the number of revolutions is mainly restricted by the opening angle $\psi$. The limiting opening angle $\psi$ allowing at least one complete revolution (i.e. $\Delta\phi \gtreqqless 2\pi$) about the jet axis—as observed for

$C4^*$ can be determined. For very large angular momentum $L$, the second *arctan* term in Equation (16) may be neglected, and for $t \to \infty$ we obtain the condition

$$\Delta\phi = 2\pi = \frac{1}{\sin\psi_{\max}} \arctan(\infty) = \frac{\pi}{2\sin\psi_{\max}}. \quad (17)$$

It follows that $\psi_{\max} \simeq 14.5°$. Thus the parameter space for C4 can be restricted to $\psi < 14.5°$, because at least one oscillation has been observed since its discovery (fig. 7).
We point out that the number of revolutions observed with VLBI is not necessarily equal to the number of revolutions made by the component since its creation. VLBI-observations do not penetrate the jet down to the region where components are created. This is supported by optical variability observations and theoretical simulations, which imply that most of the revolutions take place closer to the central engine (Schramm et al. 1993; Camenzind & Krockenberger 1992). In our analysis, we start calculations at the position of the VLBI-core of the jet.

Figure 4 shows a "typical" trajectory of this model. In contrast to Case 2 we find no self-similarity, but an asymptotic behavior for $z \to \infty$. Close to its origin the coil of the helix is strongest, i.e. the pitch angle $\epsilon$ is smallest, but it opens along the jet axis. These properties match the characteristics of the component trajectories observed in 3C 345.

### 2.4. Case 4

In this case, the quantities $p_z$, $L_z$ and $\psi$ are conserved. Again, the solutions are derived by integration (now we use Equations (3, 4 & 2):

$$z_{(t)} = v_z t \quad (18)$$
$$r_{(t)} = v_z t \tan\psi + r_0 \quad (19)$$
$$\phi_{(t)} = \phi_0 - \frac{L_z}{v_z r_{(t)} \tan\psi} + \frac{L_z}{v_z r_0 \tan\psi} \quad (20)$$

For clarity we write $\phi_{(t)}$ as a function of $r_{(t)}$, which itself is a linear function of time (eq. 19). For large $z$ the trajectory degenerates into a straight line, since the phase angle $\phi_{(t)}$ approaches a fixed value $\phi_\infty$, given by

$$\phi_\infty = \lim_{t\to\infty} \phi_{(t)} = \phi_0 + \frac{L_z}{v_z r_0 \tan\psi}. \quad (21)$$

The trajectories of this case are very similar to those in Case 3 (Fig. 4). In the following paragraph we discuss the discrimination between these two cases and te application to 3C 345.

### 3. Application to 3C 345

Of the four cases discussed above, we rejected Cases 1 and 2 as they do not reflect the basic character of the trajectories observed in 3C 345. Cases 3 and 4 can produce very similar trajectories (fig. 4), but they yield different typical velocity profiles as shown in Fig. 5. We favor Case 3 over Case 4, as it does not only give a better formal fit to the kinematic data, but unlike Case 4 it also predicts that component speeds remain on a high level as a component separates from the core.

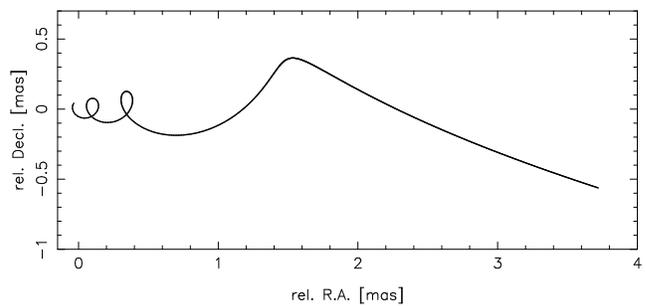

**Fig. 4.** A typical trajectory illustrating the properties of models 3 and 4. The trajectory asymptotically transforms into a straight line.

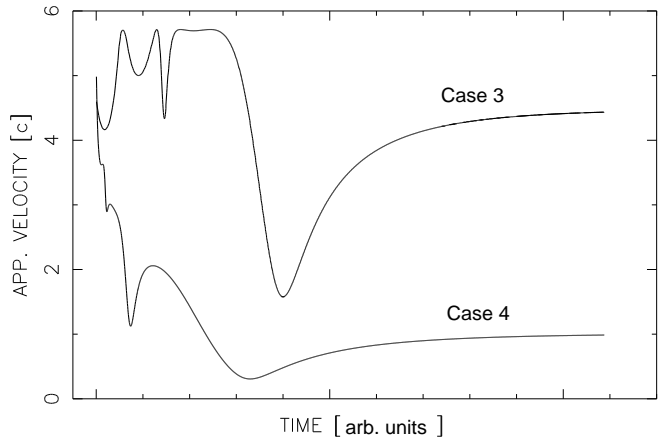

**Fig. 5.** The apparent velocities versus time in the source frame are plotted for Cases 3 and 4 corresponding to the path in Fig. 4. The apparent velocity oscillates strongly where the trajectory shows loops, later converging towards a constant value were the helix opens. For Case 4 the apparent velocity shows an overall decreasing trend due to the non-conservation of the angular momentum and the associated loss of the rotational velocity component. In an astrophysical jet this could be due to the interaction with the ambient medium.

This is in accordance with the result that component speeds in 3C 345 appear to increase with distance from the core (Zensus et al. 1995). We note, that the preference for Case 3 here does not preclude that the other model variants might be applicable in other sources.

The final set of parameters for the chosen model variant was derived by fitting to the (different) curved trajectories and velocity profiles for components C4 and C5 in 3C 345 (Fig. 7–10). For our purpose, velocities were computed from subsequent observations at a given frequency. Furthermore, we restricted the analysis to the inner 3 milliarcseconds from the core. At larger distances, the components follow a similar curved path that is indicative of large-scale bending of the jet axis not accounted for in our model at this point (cf. Unwin & Wehrle 1992, Qian et al. 1992, Zensus et al. 1995).

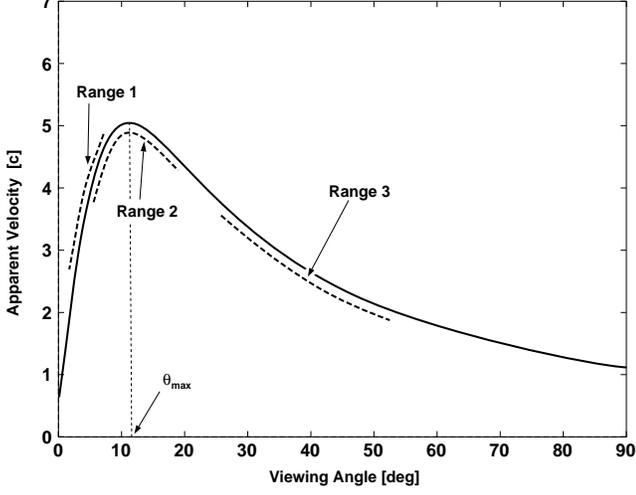

**Fig. 6.** The range covered by the angle to the line of sight determines the number of maxima during one turn around the axis on the helical path. Ranges 1 and 3 are separated by $\theta_{\max}$ where the apparent speed is at maximum.

*3.1. Constraints*

The projected trajectory on the sky plane is fully determined by eight partly-interdependent parameters, assuming that the origin of the coordinate system is coincident with the stationary VLBI-core. They correspond to the three conservation quantities kinetic energy, angular momentum, and opening-angle; to the three constants of integration; and to two parameters orienting the trajectory on the sky.

The practical model calculations are based on parameters equivalent to the above: the initial value for the Lorentz factor $\gamma$ of a component, the angular velocity $\omega_{\rm o}$, the opening angle $\psi$, the time of appearance from the VLBI-core $t_{\rm ej}$, the distance from the jet axis $r_{\rm o}$, the phase angle of the helix $\phi_{\rm o}$, the angle between the jet axis and the observer's line of sight $\theta$, and the position angle of the jet axis in the sky plane $\chi$. In addition, we allow for an offset of the observed VLBI-core from the jet axis in right ascension ($\Delta \alpha$) and in declination ($\Delta \delta$).

Several constraints follow directly from the observations for component C4. The lower limit of the Lorentz factor $\gamma$ is approximately given by the highest observed superluminal speed $v_{\max} \approx 5$ (fig. 8, $H_{\rm o} = 100\,{\rm km\,s^{-1}\,Mpc^{-1}}$, $q_{\rm o} = 0.5$). This implies that that the angle to the line of sight $\theta_{(t)}$ must be smaller than $\theta_{(t_x)} \approx 20°$ for these points, if we assume that the Lorentz factor is not larger than 100. The opening angle, inferred from the envelope of the trajectories, is $\psi_{\rm obs} \approx 13°$ (Biretta et al. 1986). Accounting for projection effects, the true opening angle is $\psi < 13°$. The angular speed of rotation around the jet axis can be estimated from the observation that C4 has undergone a complete oscillation about the axis in $t_{\rm obs} \approx 9$ years. If $\gamma \approx 5$, we obtain an estimated lower limit for the initial angular velocity, corrected for redshift and light travel-time effects (Qian et al. 1995) using $\cos\theta = \beta$:

$$\omega_{\rm o} = \frac{2\pi}{t_{\rm obs}} \frac{(1+z)}{\gamma D} = \frac{2\pi}{t_{\rm obs}} \frac{(1+z)}{\gamma^2} \gtrsim 2°\,{\rm yr}^{-1}, \qquad (22)$$

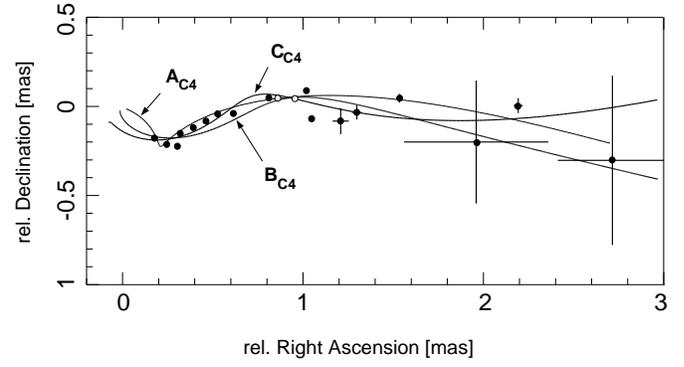

**Fig. 7.** Fits to the observed trajectory of C4 (corresponding to solutions A, B, and C in Table 1). Filled circles represent the data. Where not shown, error bars are smaller than the symbol.

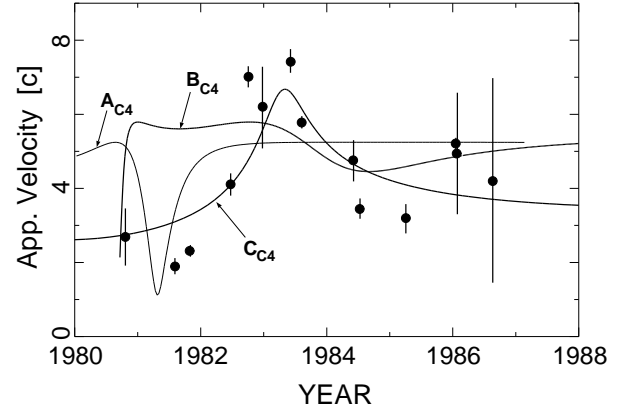

**Fig. 8.** Observed apparent speed of component C4 compared to the model calculations (see the trajectories in fig. 7).

where $\gamma = (1-\beta^2)^{-\frac{1}{2}}$ is the Lorentz factor of the bulk motion, and $D = [\gamma(1-\beta\cos\theta)]^{-1}$ is the Doppler factor.

The limit for the initial radial distance from the jet axis $r_{\rm o}$ follows from the smallest observed core-distance of C4. $t_{\rm ej}$ is estimated from the zero separation time as given by Zensus et al. (1995). In summary, we restrict the allowed parameter space for our model to the following values:

| Lorentz factor | $\gamma$ | $\gtrsim$ | 5.0 |
| opening angle | $\psi$ | $\lesssim$ | 5° |
| angular velocity | $\omega_{\rm o}$ | $\gtrsim$ | 2°/ yr |
| initial radius | $r_{\rm o}$ | $\lesssim$ | 3 ly |
| time of appearance | $t_{\rm ej}$ | $\approx$ | 1979.5 |

The observation that the outer components C3 and C2 follow almost straight trajectories (Zensus et al. 1995) is used as a further constraint.

**Table 1.** Parameters of the best fits found for model 3. As indicated in column 2, A and B are solutions of small angles to the line of sight (range 1 in Figure 6) and positive or negative sense of rotation. Solution C is a large-angle solution (range 3 in Figure 6) showing different senses of rotation for components C4 and C5.

| Solut. | Rot. | Range $\theta$ | $t_{ej}$ | $\theta$ [°] | $\gamma_o$ | $\omega_o$ [$\frac{°}{yr}$] | $r_o$ [ly] | $\psi$ [°] | $\phi_o$ [°] | $\chi$ [°] | $\Delta\delta$ [mas] | $\Delta\alpha$ [mas] |
|---|---|---|---|---|---|---|---|---|---|---|---|---|
| $A_{C4}$ | − | 1 | 1979.9 | 7.5 | 5.35 | −8.0 | 1.10 | 1.0 | 60 | 265 | −0.07 | 0.0 |
| $A_{C5}$ | − | 1 | 1982.3 | 7.5 | 5.0 | −6.6 | 1.15 | 1.15 | 140 | 265 | −0.07 | 0.0 |
| $B_{C4}$ | + | 1 | 1980.1 | 6.8 | 5.8 | +11.5 | 0.7 | 0.95 | 100 | 263 | −0.07 | −0.03 |
| $B_{C5}$ | + | 1 | 1982.3 | 6.8 | 4.6 | +9.7 | 0.8 | 0.95 | 290 | 263 | −0.07 | −0.03 |
| $C_{C4}$ | + | 3 | 1977.8 | 26 | 10.0 | +29.8 | 0.7 | 1.55 | 60 | 277 | −0.15 | −0.10 |
| $C_{C5}$ | − | 3 | 1981.0 | 27 | 4.2 | −10.9 | 1.5 | 2.50 | 80 | 260 | 0.00 | 0.00 |

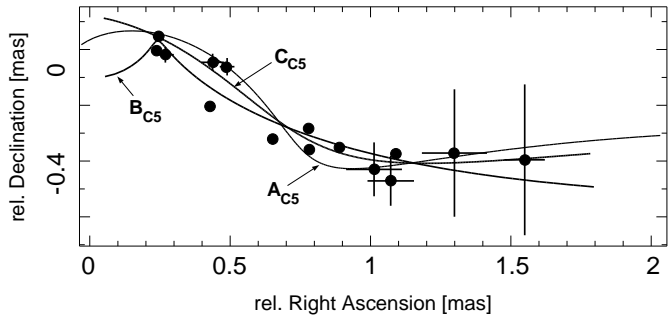

**Fig. 9.** Fits to the observed trajectory of C5 (see also Table 1).

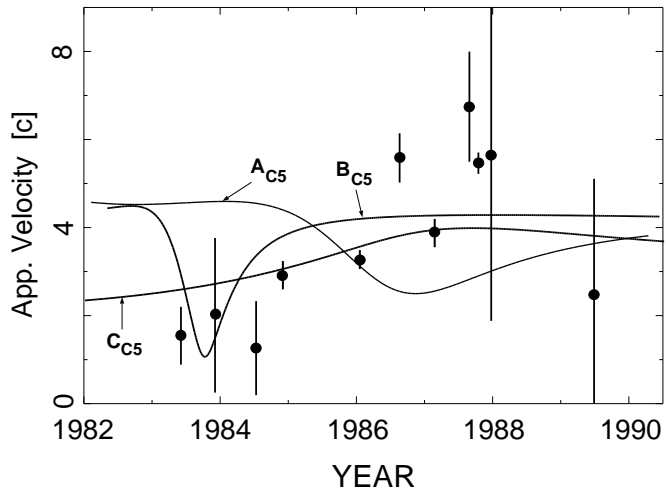

**Fig. 10.** Observed apparent speed of component C5 compared to the model calculations (see the trajectories in fig. 9).

## 4. Results

We fitted various sets of parameters and found that only few solutions are possible that are qualitatively different. They can be distinguished by their sense of rotation (given by the sign of $\omega_o$) and the range covered by the angle to the line of sight $\theta_{(t)}$ during motion along the trajectory (see the $(\theta, v_{app})$-diagram in Figure 6). With the three ranges indicated in Figure 6 and the two senses of rotation there are six qualitatively different solutions possible for every component. Of these, three have been found for every component to fit the trajectories (see Table 1).

In summary we find that :

(a) The basic motion of components C4 and C5 in 3C 345 can be explained by a helical path based on the conservation of kinetic energy and angular momentum combined with a constant opening angle of the jet (Case 3 in Section 2.3).

(b) Formally the best fit of the individual trajectories and the apparent velocities is represented by the large-angle solution C (the angle to the line of sight is larger than the angle $\theta_{max}$ of maximum apparent speed $v_{max}$ for a given Lorentz factor $\gamma$). Here components C4 and C5 have opposite senses of rotation.

(c) The trajectories of C4 and C5 could be fitted with similar sets of parameters only for small-angle solutions A and B (the angle between the jet axis and the line of sight was smaller than $\theta_{max}$ for a given Lorentz factor $\gamma$). Here components C4 and C5 have the same senses of rotation (in contrast to the large-angle solution C). We regard this as the only reasonable solution considering the physical fundamentals of our model. We cannot exclude, however, that apparent opposite senses of rotation might be observable, if the observed apparent motion reflects merely phase velocities or if solid rotation of the entire jet is superimposed on the helical motion of individual components.

(d) The fit is considerably better if the jet axis does not pass through the VLBI-core (requiring shifts in both right ascension and declination of approximately 0.07 mas).

(e) The derived Lorentz factor for C5 is always smaller than the one obtained for C4 (see Table 1). The deduced distance from the jet axis are larger for C5 than for C4. Components C6 and C7 seem to move on largely different paths than components C4 and C5 (Krichbaum et al., 1993a, Zensus et al., 1995). Such behaviour is also consistent with the view of a shell like structure or nested conical jets expected from MHD models (e. g., Camenzind, 1986).

Solution B represents the best fit found for the combined data of components C4 and C5. The main difference compared to solution A showing in the fits to the apparent velocities; around

1984, solution A has a high value when the observed apparent velocity of C5 is low, whereas the opposite is true around 1987. However, except for C4 around 1982, the basic behavior of solution B agrees with the observation for both components. Solution B yields the following physical parameters (see Table 1): Lorentz factors $\gamma = 5.8$ and $4.6$ for C4 and C5, respectively, and for both components a mean angle to the line of sight $\theta = 6.8°$, using a opening angle $\psi$ of $0.95°$. These numbers are very close to the values $\gamma = 7.2 \pm 1.0$ and $\theta = 6.8° \pm 1.5°$ obtained by Unwin & Wehrle (1992) and also to those deduced by Zensus et al. (1995). They find a lower limit for the Lorentz factor $\gamma \geq 8$ and the angle to the line of sight $\theta \leq 5.4°$.

### 4.1. Error estimates

Fixing all other parameters, the possible ranges for the Lorentz factor $\gamma$, for the angle to the line of sight $\theta$, and for the year of ejection $t_{ej}$ are about 10 percent. The initial phase angle $\phi_o$ of the helix is determined to within $20°$, whereas the position angle $\chi$ can only vary by about $2°$ to yield a fit consistent with the data to about $1\sigma$. The initial angular velocity $\omega_o$ and the radius $r_o$ are fixed to within less than $0.5 \deg \mathrm{yr}^{-1}$ and $0.1$ lightyears, respectively. However, the uncertainty is several times larger, if one allows $\omega_o$ and $r_o$ to vary simultaneously. Another strong dependency exists between $\gamma$ and $\theta$. $\gamma$ may vary by about $\pm 2$, and $\theta$ about $\pm 3°$. The opening angle $\psi$ may vary within a few tenths of a degree. For solution C the error ranges probably are larger by an estimated factor of 3. In general, the error ranges depend on the range covered by the solution in the $(\theta, v_{app})$-diagram (Fig. 6).

### 5. Conclusion

Our analysis demonstrates that the basic properties of component trajectories and the apparent velocities for components C4 and C5 in 3C 345 can be described by a simple scenario of helical motion with a straight jet axis, based on conservation of physical quantities. However, there are likely to be additional small-scale effects at work which alter the smooth behaviour predicted by the model. These effects must be sufficiently large to prevent a unique solution that would explain the complete kinematic behaviour of even one single component. Similarly to the approach by Qian et al. (1992), it was not possible to establish a unique sense of rotation about the jet axis. Extension of the model to include a curved overall jet axis, and detailed monitoring of new components, especially closer than a milliarcsecond to the core, might eventually allow such a unique solution.

Additional constraints should be obtained from the observed flux density evolution of the components at different frequencies. For example, from the spectral evolution the nature of the emission regions can be clarified (Lobanov & Zensus 1994, 1995), discriminating geometric and kinematic scenarios. We still do not know if the moving emission regions are, e. g., turbulent plasma inhomogeneities or shock waves, which might be decided from synchrotron radiation transfer calculations (cf. Gómez et al. 1993). Since motion of a specific component is typically not completely regular, turbulent motion also might play a role causing deviations from the smooth paths predicted by our model.

We conclude that hydromagnetic models without complex, turbulence-like motion can only account for the smooth overall trajectory of individual emission regions. Closer to the core (i. e. within less than one milliarcsecond in the case of 3C 345), such smooth trajectories should prevail, as there the ordered magnetic fields associated with the central engine should be dominating the source kinematics.

*Acknowledgements.* We thank M. Camenzind, K.-H. Wüllner, C. Rabaça, A. Lobanov, and H. Lesch, for valuable discussions. We acknowledge support by the *Studienstiftung des Deutschen Volkes* (WS) and the *BMFT Verbundforschung* (TPK). The National Radio Astronomy Observatory is operated by Associated Universities, Inc., under cooperative agreement with the National Science Foundation.